\documentclass[11pt, oneside]{article}  

\usepackage{geometry}  
\usepackage{cmbright}
\usepackage{graphicx}				
\usepackage{tcolorbox}
\usepackage{amssymb}
\usepackage{longtable}
\usepackage{amsmath}
\usepackage{booktabs}
\usepackage{url}
\usepackage{color}
\definecolor{c1}{rgb}{0.12, 0.56, 1.0}
\usepackage{xspace}
\usepackage{xcolor}
\usepackage{caption}
\usepackage{subcaption}
\usepackage{authblk} 

\usepackage{sectsty} 
\definecolor{cool_blue}{RGB}{24, 132, 193}
\sectionfont{\color{c1}\selectfont}
\subsectionfont{\color{c1}\selectfont}
\subsubsectionfont{\color{c1}\selectfont}
\subparagraphfont{\color{gray}\selectfont}

\definecolor{fruitpushorange}{RGB}{255, 127, 0}
\newcommand{\data}{$(s_1, ..., s_k)$\xspace}

\usepackage{soul}

\title{A Bayesian treatment of the German tank problem}
\author[1]{Cory M. Simon}
\affil[1]{School of Chemical, Biological, and Environmental Engineering. Oregon State University. Corvallis, OR. USA. }
\affil[]{\texttt{cory.simon@oregonstate.edu}}

\begin{document}
\maketitle

\begin{abstract}
	The German tank problem has an interesting historical background and is an engaging problem of statistical estimation for the classroom. The objective is to estimate the size of a population of tanks inscribed with sequential serial numbers, from a random sample. In this tutorial article, we outline the Bayesian approach to the German tank problem, (i) whose solution assigns a probability to each tank population size, thereby quantifying uncertainty, and (ii) which provides an opportunity to incorporate prior information and/or beliefs about the tank population size into the solution. We illustrate with an example. Finally, we survey problems in other contexts that resemble the German tank problem.
\end{abstract}

\begin{figure}[h!]
	\centering
 \includegraphics[width=0.4\textwidth]{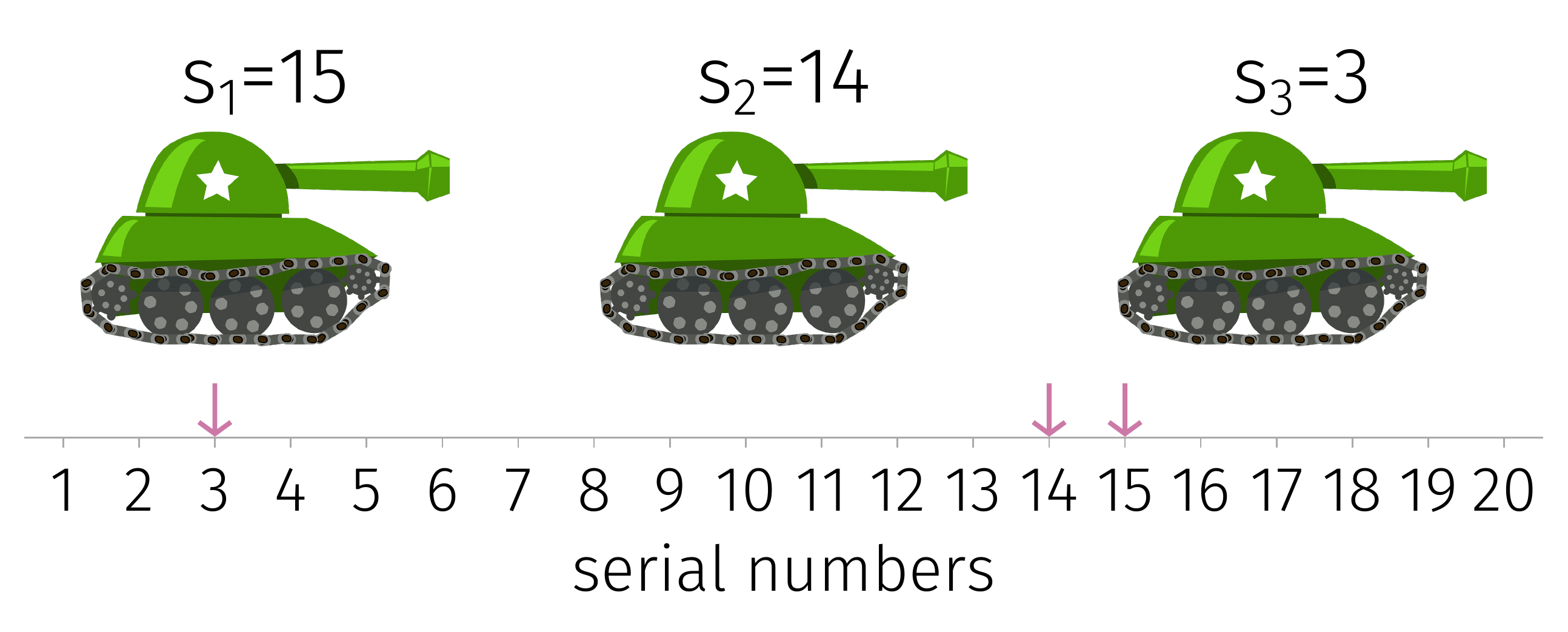}
 \includegraphics[width=0.4\textwidth]{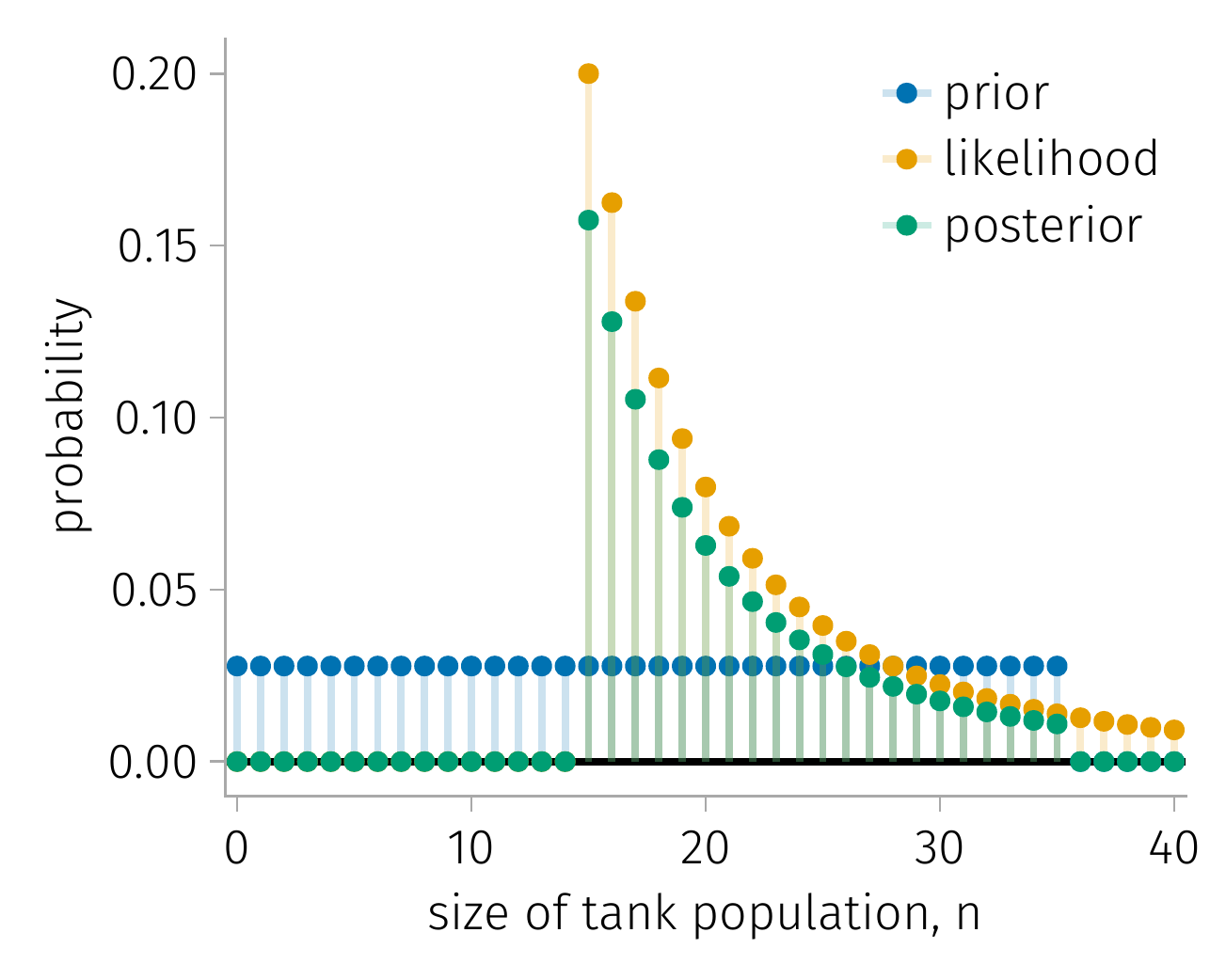}
\end{figure}

\clearpage

\section{Background}

\subsection{History}
To inform their military strategy during World War II (1939-1945), the Allies sought to estimate Germany's rate of production of various military equipment (tanks, tires, rockets, etc.).
Conventional methods to estimate armament production---including 
(i) extrapolating data on prewar manufacturing capabilities, 
(ii) obtaining reports from secret sources, and 
(iii) interrogating prisoners of war---were unreliable and/or contradictory. 

In 1943, British and American economic intelligence agencies exploited a German manufacturing practice in order to statistically estimate their armament production. 
Specifically, Germany marked their military equipment with serial numbers and codes for the date and/or place of manufacture. Their intention was to facilitate handling spare parts and trace defective equipment/parts back to the manufacturer for quality control.
However, these markings on a captured sample of German equipment conveyed information to the Allies about Germany's production of it.

To estimate Germany's production of tanks, the Allies collected serial numbers on the chassis, engines, gearboxes, and bogie wheels of samples of tanks by inspecting captured tanks and examining captured records\footnote{Eg., captured records from tank repair depots listed serial numbers of the chassis and engine of repaired tanks, and records from divisional headquarters listed chassis serial numbers of tanks held by a specific unit.}. 
Despite lacking an \emph{exhaustive} sample, the sequential nature of\footnote{
Gearboxes on captured tanks, for example, were inscribed with serial numbers belonging to an unbroken sequence. Chassis serial numbers, on the other hand, were broken into blocks to distinguish models/designs, leaving gaps between the serial numbers assigned to them.
} and patterns in these samples of serial numbers enabled the Allies to estimate Germany's tank production---postwar, we know---much more accurately than conventional intelligence methods (Tab.~\ref{tab:success}).

See Ruggles and Brodie \cite{ruggles1947empirical} for the detailed historical account of serial number analysis to estimate German armament production during World War II.

\begin{table}[h!]
\centering 
\caption{Monthly production of tanks by Germany. \cite{ruggles1947empirical}} \label{tab:success}
\begin{tabular}{p{2.5cm} p{4cm} p{4cm} p{2cm}}
\toprule
 & \multicolumn{2}{c}{estimates} &   \\ 
\cmidrule(r){2-3}
date & conventional American \& British Intelligence & serial number analysis  & German records \\
\midrule
June, 1940 & 1000& 169 &  122 \\
June, 1941 &1550 & 244 &   271 \\
August, 1942 & 1550& 327  & 342 \\
\bottomrule
\end{tabular}
\end{table}

\subsection{The German tank problem}
Simplification of the historical context to estimate German tank production via serial number analysis \cite{ruggles1947empirical} motivated the formulation of the textbook-friendly \emph{German tank problem} \cite{goodman1952serial}: 
\begin{tcolorbox}[title=Problem statement, colback=white, colframe=c1]
\noindent 
In the backdrop of World War II, the German military has $n$ tanks. 
Each tank is inscribed with a unique serial number in $\{1, ..., n\}$. \\

As the Allies, we do not know $n$, but we captured (without replacement, of course) a sample of $k$ German tanks with inscribed serial numbers \data. 

\begin{center}
	\begin{tabular}{cccc}
		\includegraphics[width=0.125\textwidth]{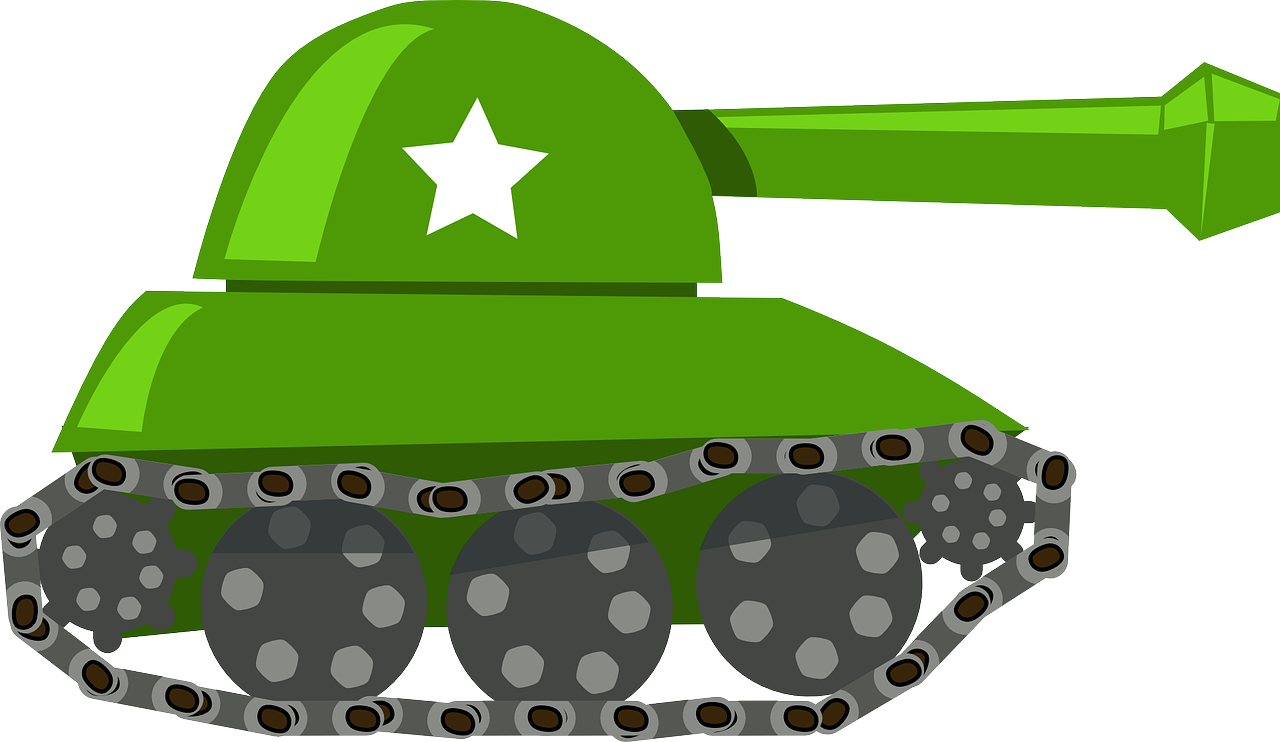} &  \includegraphics[width=0.125\textwidth]{tank.png}  & & \includegraphics[width=0.125\textwidth]{tank.png} \\
		\large $s_1$ & \large $s_2$ & \large $\cdots$ & \large $s_k$ 
	\end{tabular}
\end{center}

Assuming all tanks in the population were equally likely to be captured, our objective is to estimate $n$ in consideration of the data \data.
\end{tcolorbox}

In 1942, Alan Turing and Andrew Gleason discussed a variant of the German tank problem, ``how to best to estimate the total number of taxicabs in a town, having seen a random selection of their license numbers'', in a crowded restaurant in Washington DC \cite{hodges2014alan,hall2014alan}. 
Today, with its interesting historical background \cite{ruggles1947empirical}, 
the German tank problem is still a suitable conversation topic for dinners and serves as an intellectually engaging, challenging, and enjoyable problem to illustrate combinatorics and statistical estimation in the classroom \cite{johnson1994estimating,berg2021bayesian,mosteller1987fifty,downey2021think}.

\paragraph{Uncertainty quantification.}
Any estimate of the tank population size $n$ from the data \data is subject to uncertainty, since we (presumably) have not captured \emph{all} of the tanks (ie., $k\neq n$, probably).  
Quantifying uncertainty in our estimate of $n$ is important because high-stakes military decisions may be made on its basis.

\paragraph{Our contribution.}
In this pedagogical article, we outline the Bayesian approach to the German tank problem, 
(i) whose solution assigns a probability to each tank population size, thereby quantifying uncertainty, and
(ii) which provides an opportunity to incorporate prior information and/or beliefs about the tank population size into the solution.

\subsection{Survey of previous work on the German tank problem}
\paragraph{The frequentist approach.}
Border \cite{bordernotes} calls the German tank problem a "weird case" in frequentist estimation. The maximum likelihood estimator of the tank population size $n$ is the maximum serial number observed among the $k$ captured tanks, $m^{(k)}:=\max_{i \in \{1, ..., k\}} s_i$. This is a biased estimator, as certainly $m^{(k)} \leq n$.

Goodman \cite{goodman1952serial,goodman1954some} derives the minimum-variance, unbiased estimator of the tank population size
\begin{equation}
	\hat{n} = m^{(k)} + \left(\frac{m^{(k)}}{k}-1 \right). \label{eq:nhat}
\end{equation}
To intuit $\hat{n}$, note (i) $n$ must be greater than or equal to $m^{(k)}$ and (ii) if we observe large (small) gaps between the serial numbers \data after sorting them (incl.\ the gap preceding the smallest serial number), then $n$ is likely (unlikely) to be much greater than $m^{(k)}$. 
The estimator of $n$ in eqn.~\ref{eq:nhat} quantifies how far beyond the maximum serial number $m^{(k)}$ we should estimate the tank population size, based on the gaps; $m^{(k)}/k-1$ is the average size of the gaps.
Goodman also derives a frequentist confidence interval for $n$.

Clark, Gonye, and Miller explore using simulations and linear regression to discover the estimator in eqn.~\ref{eq:nhat} \cite{clark2021lessons}. 

\paragraph{For pedagogy.} Champkin highlights the historical context of the German tank problem as a "great moment in statistics" \cite{grajalez2013great}. 
Johnson lists and evaluates several intuitive point estimators for the size of the tank population \cite{johnson1994estimating}. 
Scheaffer, Watkins, Gnanadesikan, and Witmer \cite{scheaffer2013activity} propose a hands-on learning activity to illustrate the German tank problem by sampling chips, labeled with numbers from 1 to $n$, from a bowl. 
Berg \cite{berg2021bayesian} uses the German tank problem as a competition in the classroom.  

\paragraph{The Bayesian approach.}
Closely related to our paper, Roberts \cite{roberts1967informative}, H{\"o}hle and Held \cite{hohle2006bayesian}, and Linden, Dose, and Toussaint \cite{von2014bayesian}, and Cocco, Monasson, and Zamponi \cite{cocco2022statistical} provide a Bayesian analysis of the German tank problem. They derive an analytical formula for the mean of the posterior distribution of the tank population size under an improper, uniform prior distribution. Andrews \cite{blogpost} outlines the Bayesian approach to the German tank problem in a blog post containing code in the R language.

\paragraph{Generalizations/variants.}
Goodman \cite{goodman1952serial,goodman1954some} poses a variant of the German tank problem where the initial serial number is not known; ie., where the $n$ tanks are inscribed with serial numbers $\{b+1, ..., n+b\}$ with $b$ and $n$ unknown. 
Lee and Miller generalize the German tank problem to the settings where the serial numbers are continuous and/or lie in two dimensions \cite{lee2022generalizing}. 

\subsection{Overview of the Bayesian approach to the German tank problem}
Under a Bayesian perspective \cite{bolstad2016introduction,downey2021think,van2021bayesian}, we treat the (unknown) total number of tanks as a discrete random variable $N$ (hence, capitalization) to model our uncertainty in it. A probability mass function of $N$ assigns a probability to each possible tank population size $n$. This probability is a measure of our degree of belief, perhaps with some basis in knowledge/data, that the tank population size is $n$ \cite{ghosh2006introduction}. 

Because the observed serial numbers \data provide information about the tank population size, the probability mass function of $N$ differs before and after they are collected and considered. Hence, $N$ has a \emph{prior} and \emph{posterior} probability mass function. 

The three inputs to a Bayesian treatment of the German tank problem are: 
\begin{enumerate}
	\item the \emph{prior} mass function of $N$, which expresses a combination of our subjective beliefs and objective knowledge about the tank population size before we collect and consider the sample of serial numbers.
	\item the \emph{data}, the observed serial numbers \data, viewed as realizations of random variables owing to the stochasticity of tank-capturing.
	\item the \emph{likelihood} function, giving the probability of the data \data under each tank population size $N=n$, based on a probabilistic model of the tank-capturing process.
\end{enumerate}

The output of a Bayesian treatment of the German tank problem is the \emph{posterior} mass function of the tank population size $N$, conditioned on the data \data. The posterior follows from Bayes' theorem and can be viewed as an update to the prior in light of the data.
The posterior mass function of $N$ assigns each possible tank population size $n$ with a probability according to a compromise between its
(i) likelihood, which quantifies the support the observed serial numbers \data lend to the tank population size being $n$ according to our probabilistic tank-capturing model, 
and 
(ii) prior probability, which quantifies how likely we thought the tank population size might be $n$ before the serial numbers \data were collected and considered. \cite{van2021bayesian}

The posterior mass function of $N$ is the raw, uncertainty-quantifying, Bayesian solution to the German tank problem. We may summarize the posterior by reporting its median and the high-mass subset of the natural numbers that credibly contains the tank population size. Also, we can use the posterior to answer questions such as, "what is the probability that $N$ exceeds some threshold quantity $n^\prime$ that would alter military strategy?".


\section{A Bayesian approach to the German tank problem}
We now tackle the German tank problem from a Bayesian standpoint. 

For reference, the variables are listed in Tab.~\ref{tab:params}. 
We use upper- and lower-case letters to represent random variables and realizations of them, respectively. 
Throughout, we employ the indicator function $\mathcal{I}_{A}(x)$ which maps its input $x$ to 1 if $x$ belongs to the set $A$ and to 0 otherwise (if $x\notin A$).

\begin{table}[h!]
	\centering
	\caption{List of parameters/variables.} \label{tab:params}
	\begin{tabular}{c c l}
		\toprule
		parameter/variable & $\in$ & description \\
		\midrule
		$n$ & $\mathbb{N}_{\geq 0}$ & size of population of tanks \\
		$k$ & $\mathbb{N}_{>0}$ &  number of captured tanks \\
		$s_i$ & $\mathbb{N}_{>0}$ &  serial number on captured tank $i$ \\
		$s^{(k)}$ & $\mathbb{N}_{>0}^k$ & vector listing the serial numbers on the $k$ captured tanks \\
		$m^{(k)}$ & $\mathbb{N}_{>0}$ &  maximum serial number among the $k$ captured tanks \\
		\bottomrule
	\end{tabular}
\end{table}

\subsection{The data, data-generating process, and likelihood function}
\paragraph{The data.} The data we obtain in the German tank problem is the vector of serial numbers inscribed on the $k$ captured tanks
\begin{equation}
	s^{(k)}:=(s_1,...,s_k).
\end{equation} 
We view the data $s^{(k)}$ as a realization of the discrete random vector $S^{(k)}:=(S_1, ..., S_k)$. Note, at this point, we are entertaining the possibility that the order in which tanks are captured matters.

\paragraph{The data-generating process.}
The stochastic \emph{data-generating process} constitutes sequential capture of $k$ tanks from a population of $n$ tanks, without replacement, then inspecting their serial numbers to construct $s^{(k)}$.
We assume that each tank in the population is equally likely to be captured at each step.
Then, mathematically, the stochastic data-generating process is sequential, uniform random selection of $k$ integers, without replacement, from the set $\{1, ..., n\}$.

\paragraph{The likelihood function.}
The \emph{likelihood function} specifies the probability of the data $S^{(k)}=s^{(k)}$ given each tank population size $N=n$.
Each outcome $s^{(k)}$ in the sample space $\Omega_n^{(k)}$ is equally likely, where
\begin{equation}
	\Omega_n^{(k)} := \{ (s_1, ..., s_k)_{\neq}  :  s_i \in \{1, ..., n\} \; \text{for all } i \in \{ 1,..., k \} \},
\end{equation} with $(\cdots)_{\neq}$ meaning the elements of the vector $(\cdots)$ are unique. 
The number of outcomes in the sample space, $|\Omega_n^{(k)}|$, is the number of distinct ordered arrangements of $k$ distinct integers from the set $\{1,...,n\}$, given by the falling factorial:
\begin{equation}
	(n)_k:= n(n-1)\cdots (n-k+1) = n! / (n-k)!.
\end{equation}
Under the data-generating process, then, the probability of observing data $S^{(k)}=s^{(k)}$ given the tank population size $N=n$ is the uniform distribution:
\begin{equation}
	\pi_{\text{likelihood}}(S^{(k)}=s^{(k)} \mid N=n)=
	\dfrac{1}{(n)_k} \mathcal{I}_{\Omega_n^{(k)}}\left(s^{(k)}\right).
	\label{eq:dgp}
\end{equation}

\subparagraph{Interpretation.}
The likelihood quantifies the support the serial numbers on the $k$ captured tanks in $s^{(k)}$ lend for any particular tank population size $n$, according to our probabilistic model of the tank-capturing process \cite{van2021bayesian}. We view $\pi_{\text{likelihood}}(S^{(k)}=s^{(k)} \mid N=n)$ as a function of $n$, since in practice we possess the data $s^{(k)}$ but not $n$.

\subparagraph{The likelihood as a sequence of events.} 
Alternatively, we may arrive at eqn.~\ref{eq:dgp} from a perspective of sequential events $S_1=s_1, S_2=s_2, ..., S_k=s_k$. First, the probability of a given serial number on the $i$th captured tank, conditioned on the tank population size and the outcomes of the previous serial numbers, is the uniform distribution
\begin{equation}
	\pi (S_i=s_i \mid N=n, S_1=s_1, ..., S_{i-1}=s_{i-1})=\dfrac{1}{n-i+1} \mathcal{I}_{ \{1,...,n\} \setminus \{s_1, ..., s_{i-1}\}}(s_i)
\end{equation}
since there are $n-i+1$ tanks to choose from at uniform random.
By the chain rule, the joint probability
\begin{equation}
	\pi_{\text{likelihood}}(S_1=s_1, ..., S_k=s_k \mid N=n) = \displaystyle \prod_{i=1}^k \pi (S_i=s_i \mid N=n, S_1=s_1,...,S_{i-1}=s_{i-1})
\end{equation}
giving eqn.~\ref{eq:dgp} after simplifying the product of indicator functions.

\subparagraph{The likelihood function in terms of the maximum observed serial number.}
We will find in Sec.~\ref{sec:posterior} that only two independent features of the data \data provide information about the tank population size, $N$: its (i) size, $k$, and (ii) maximum observed serial number 
\begin{equation}
    m^{(k)} = \max_{i \in \{1, ..., k\}} s_i .
\end{equation} 
Thus, we also write a different likelihood: the probability of observing a maximum serial number $m^{(k)}$ given the tank population size $N=n$,  $\pi_{\text{likelihood}}(M^{(k)}=m^{(k)} \mid N=n)$.

Because each outcome $s^{(k)}\in \Omega_n^{(k)}$ is equally likely, $\pi_{\text{likelihood}}(M^{(k)}=m^{(k)} \mid N=n)$ is the fraction of sample space under population size $n$ where the maximum serial number is $m^{(k)}$.
To count the outcomes $(s_1, ..., s_k)\in\Omega_n^{(k)}$ where the maximum serial number is $m^{(k)}$, consider (i) one of the $k$ captured tanks has serial number $m^{(k)}$ and (ii) the remaining $k-1$ tanks have a serial number in $\{1, ..., m^{(k)}-1\}$.
For each of the $k$ possible positions of the maximum serial number in the vector $s^{(k)}$, there are $(m^{(k)}-1)_{k-1}$ distinct outcomes specifying the other $k-1$ entries.
Thus:
\begin{equation}
	\pi_{\text{likelihood}}(M^{(k)}=m^{(k)} \mid N=n)=
	\dfrac{k(m^{(k)}-1)_{k-1}}{(n)_k} \mathcal{I}_{\{k,...,n\}}(m^{(k)}). \label{eq:likelihood_m}
 \end{equation}
 
\subsection{The prior distribution}
The \emph{prior probability mass function} $\pi_{\text{prior}}(N=n)$ expresses a combination of our subjective beliefs and objective knowledge about the total number of tanks $N$ before the data \data are collected and considered. 

The prior mass function we impose on $N$ is context-dependent. 
Based on the amount of uncertainty it admits about the tank population size (measured by eg.\ entropy \cite{murphy2022probabilistic}),
prior distributions roughly belong to one of the ordinal categories of informative, weakly informative, or diffuse \cite{van2021bayesian}.
If we do not possess prior information about the tank population size, we adopt the principle of indifference and impose a diffuse prior, eg.\ a uniform distribution over a set of feasible tank population sizes. 
On the other hand, an informative prior might concentrate its mass around some estimate of the total number of tanks obtained through other means. 

Thinking ahead, about the posterior mass function of $N$, which balances the prior and the likelihood (the latter based on the data): 
(1) an informative prior will have a larger impact on the posterior than a diffuse one \cite{van2021bayesian}, which "lets the data speak for itself" \cite{downey2021think};
(2) generally, as the number of captured tanks $k$ increases (decreases), we expect the prior to have a smaller (larger) impact on the posterior \cite{downey2021think} as the data "overwhelms" the prior.

\subsection{The posterior distribution}
\label{sec:posterior}
The \emph{posterior probability mass function} of $N$ assigns a probability to each possible tank population size $n$ in consideration of its consistency with (1) the data \data, according to the likelihood in eqn.~\ref{eq:dgp}, and (2) our prior beliefs/knowledge encoded in $\pi_{\text{prior}}(N=n)$. 

The posterior distribution is a conditional distribution related to the likelihood and prior mass functions by Bayes' theorem:
\begin{equation}
	\pi_{\text{posterior}}(N=n \mid S^{(k)}=s^{(k)}) = 
	\frac{\pi_{\text{likelihood}}(S^{(k)}=s^{(k)} \mid N=n) \pi_{\text{prior}}(N=n)}{\pi_{\text{data}}(S^{(k)}=s^{(k)})}, \label{eq:post}
\end{equation} 
where the denominator is the probability of the data $s^{(k)}$:
\begin{equation}
	\pi_{\text{data}}(S^{(k)}=s^{(k)})= \displaystyle \sum_{n^\prime=0}^\infty \pi_{\text{likelihood}}(S^{(k)}=s^{(k)} \mid N=n^\prime) \pi_{\text{prior}}(N=n^\prime). \label{eq:prob_data}
\end{equation}
We view $\pi_{\text{posterior}}(N=n \mid S^{(k)}=s^{(k)})$ as a probability mass function of $N$, since in practice we have $s^{(k)}$. Then, $\pi_{\text{data}}(S^{(k)}=s^{(k)})$ is just a normalizing factor for the numerator in eqn.~\ref{eq:post}. 

Interpreting eqn.~\ref{eq:post}, the prior mass function of $N$ is \emph{updated}, in light of the data \data, to yield the posterior mass function of $N$. The posterior probability of $N=n$ is proportional to the product of the likelihood at and prior probability of $N=n$, a compromise between the likelihood and prior.

We simplify the posterior mass function of $N$ in eqn.~\ref{eq:post} by (i) substituting eqn.~\ref{eq:dgp}, (ii) restricting the sum in eqn.~\ref{eq:prob_data} to tank population sizes where the likelihood is nonzero, and (iii) noting the only two features of the data \data that appear are (a) its size $k$ and (b) the maximum serial number $m^{(k)}$:
\begin{equation}
	\pi_{\text{posterior}}(N=n \mid M^{(k)}=m^{(k)}) = 
	\frac{
		\displaystyle (n)_k^{-1} \pi_{\text{prior}}(N=n)
	}{
		\displaystyle \sum_{n^\prime=m^{(k)}}^\infty (n^\prime)_{k}^{-1}  \pi_{\text{prior}}(N=n^\prime)
	}
	\mathcal{I}_{\{m^{(k)}, m^{(k)}+1,...\}}(n)
	\label{eq:post_simple}
\end{equation}
Note, we may arrive at eqn.~\ref{eq:post_simple} through eqn.~\ref{eq:likelihood_m} as well.

\paragraph{Interpretation.}
The posterior probability mass function of $N$ in eqn.~\ref{eq:post_simple} is our raw, uncertainty-quantifying solution to the German tank problem. It assigns a probability to each tank population size $n$ in consideration of the serial numbers \data observed on the captured tanks, our probabilistic model of the tank-capturing process, and our prior beliefs and knowledge about the tank population size expressed in the prior mass function.


\paragraph{A remark on "uncertainty".}
The spread of the posterior mass function of $N$ in eqn.~\ref{eq:post_simple} reflects epistemic \cite{fox2011distinguishing} uncertainty about the tank population size.
The source of this posterior uncertainty is a lack of complete data: we have not captured all of the tanks\footnote{
Certainly, $k<n$ if there are gaps in the observed serial numbers \data. Even if there are no gaps in \data, we cannot be certain we have captured the tank with the largest serial number.
} and observed their serial numbers to be certain of the tank population size.
In practice, an additional source of posterior uncertainty about the tank population size is the possible inadequacy of the model of the tank-capturing process (uniform sampling) in eqn.~\ref{eq:dgp}. Ie., selection bias could be present in the tank-capturing process. Our analysis here neglects this source of uncertainty. 

\paragraph{Summarizing the posterior mass function of $N$.}
We may summarize the posterior mass function of $N$ with a point estimate of the tank population size and a credible subset of the natural numbers that likely\footnote{Well, "likely", under our assumptions embedded in the likelihood and prior mass functions.} contains it. 
A suitable point estimate of the tank population size is a median of the posterior mass function of $N$; by definition, the posterior probability that the tank population size is greater (less) than or equal to a median is at least 0.5.
A suitable credible subset, which entertains multiple tank population sizes, is the $\alpha$-high-mass subset \cite{hyndman1996computing}
\begin{equation}
	\mathcal{H}_\alpha := \{n^\prime : \pi_{\text{posterior}}(N=n^\prime \mid M^{(k)}=m^{(k)}) \geq \pi_\alpha\}
\end{equation} where $\pi_\alpha$ is the largest mass to satisfy 
\begin{equation}
	\pi_{\text{posterior}}(N \in \mathcal{H}_\alpha \mid M^{(k)}=m^{(k)}) \geq 1 - \alpha.
\end{equation}
In words, the $\alpha$-high-mass subset $\mathcal{H}_\alpha$ is the smallest to (i) contain at least a fraction $1-\alpha$ of the posterior mass of $N$ and (ii) ensure every tank population size belonging to it is more probable than any outside of it.

\paragraph{Querying the posterior distribution.} We may find the posterior probability that the tank population size belongs to any set of interest by summing the posterior mass over it. Eg., the probability the tank population size exceeds some number $n^\prime$ is:
\begin{equation}
	\pi_{\text{posterior}}(N> n^\prime \mid M^{(k)}=m^{(k)}) = \sum_{n=n^\prime+1}^\infty \pi_{\text{posterior}}(N=n \mid M^{(k)}=m^{(k)}).
\end{equation}

\subsubsection{Posterior predictive checking}
We may check the consistency of the data $s^{(k)}$ with the posterior mass function of $N$.
Conceptually, we can simulate new data $\tilde{s}^{(k)}$ using the model of the tank-capturing process under a sample of the tank population size from the posterior, then compare the simulated data $\tilde{s}^{(k)}$ to the real data $s^{(k)}$ \cite{https://doi.org/10.1111/rssa.12378,van2021bayesian}. 
More appropriately, we can compare the serial numbers in the real data \data with the mass function giving the probability that the tank with serial number $\tilde{s}$ would be captured under this process:
\begin{equation}
	\pi(\tilde{s} \in \tilde{S}^{(k)}) = \sum_{n^\prime=k}^\infty \frac{k}{n^\prime}\pi_{\text{posterior}}(N = n^\prime \mid S^{(k)}=s^{(k)}) \mathcal{I}_{\{1,...,n^\prime\}}(\tilde{s}), \label{eq:posterior_check}
\end{equation} since $k/n^\prime$ is the probability any given viable serial number $\tilde{s}$ will be observed given the tank population size $N=n^\prime$.


\section{Example}
We illustrate the Bayesian approach to the German tank problem through an example.

\begin{figure}[h!]
	\centering
	\begin{subfigure}[b]{0.395\textwidth}
		\includegraphics[width=\textwidth]{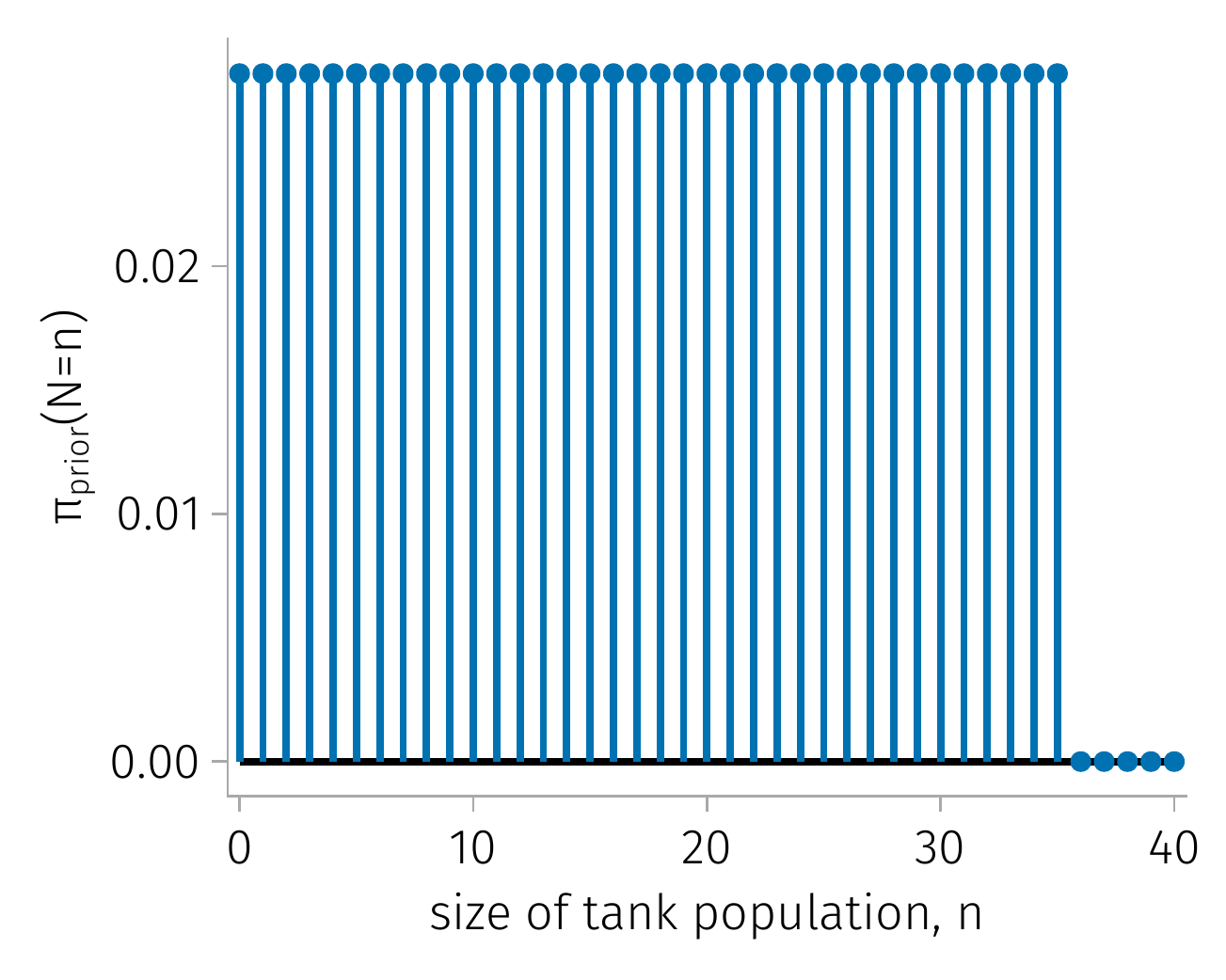} \caption{prior mass function of $N$} \label{fig:the_prior}
	\end{subfigure} 
	
	\begin{subfigure}[b]{0.5\textwidth}
		\includegraphics[width=\textwidth]{the_sample.pdf} 
	\caption{the data $s^{(k=3)}$} \label{fig:the_data}
	\end{subfigure}
	\begin{subfigure}[b]{0.395\textwidth}
		\includegraphics[width=\textwidth]{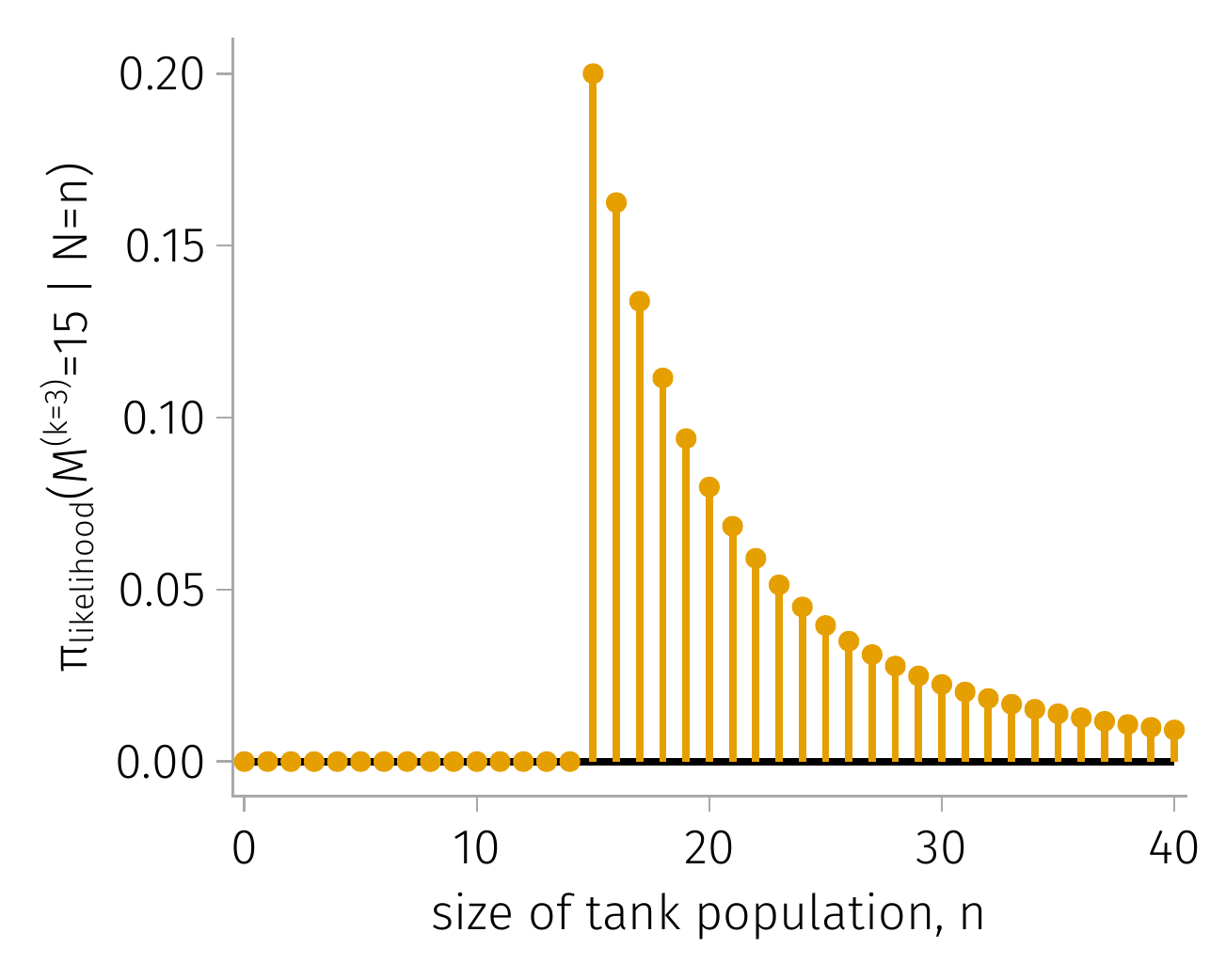} \caption{the likelihood function} 		
		\label{fig:the_likelihood}
	\end{subfigure}

	\begin{subfigure}[b]{0.395\textwidth}
		\includegraphics[width=\textwidth]{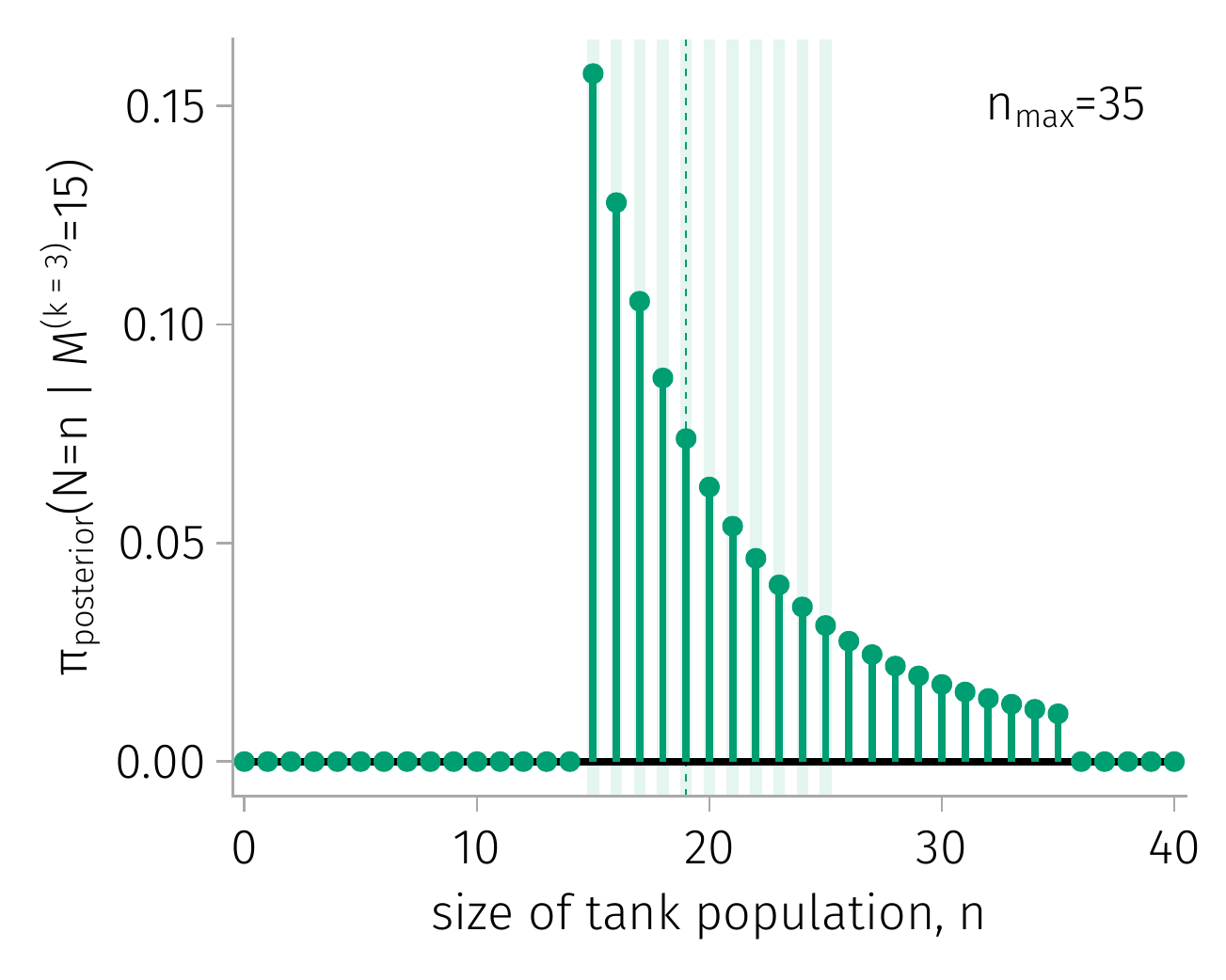} \caption{posterior mass function of $N$} \label{fig:the_posterior}
	\end{subfigure}
	\caption{A Bayesian approach to the German tank problem. 
	(a, prior) The prior mass function.
	(b, data) The data $s^{(3)}$, with maximum observed serial number $m^{(3)}=15$.
	(c, likelihood) The likelihood function associated with the data $s^{(3)}$.
	(d, posterior) The posterior mass function of $N$. $\mathcal{H}_{0.2}$ highlighted; median marked with vertical, dashed line.
	}
\end{figure}

\paragraph{The prior probability mass function of $N$.}
Suppose we have an upper bound $n_{\text{max}}$ for the possible number of tanks but no other information. Then, we may impose a diffuse prior, a uniform prior probability mass function:
\begin{equation}
	\pi_{\text{prior}}(N=n) = \dfrac{1}{n_{\max}+1} \mathcal{I}_{ \{0, ..., n_{\text{max}}\}}(n).	 \label{eq:prior}
\end{equation} 
This prior mass function expresses: in the absence of any data \data (ie., no serial numbers, not $k$ either), we believe the total number of tanks $N$ is equally likely to be a value in $\{0, ..., n_{\text{max}}\}$. Particularly, suppose $n_\text{max}=35$. Fig.~\ref{fig:the_prior} visualizes $\pi_{\text{prior}}(N=n)$.

\paragraph{The data \data.} Now suppose we capture $k=3$ tanks, with serial numbers $s^{(3)}=(15, 14, 3)$. See Fig.~\ref{fig:the_data}. So, the maximum observed serial number is $m^{(3)}=15$.

\paragraph{The posterior probability mass function of $N$.}
Under the uniform prior in eqn.~\ref{eq:prior}, the posterior probability mass function of $N$ in eqn.~\ref{eq:post_simple} becomes:
\begin{equation}
	\pi_{\text{posterior}}(N=n \mid M^{(k)}=m^{(k)})= 
		\dfrac{
			(n)_k^{-1}
			}{
			\displaystyle \sum_{ n^\prime = m^{(k)}}^{n_{\text{max}}} (n^\prime)_k^{-1}
		} \mathcal{I}_{ \{m^{(k)}, m^{(k)}+1, ..., n_\text{max}\} }(n).
\end{equation}
Fig.~\ref{fig:the_posterior} visualizes the posterior probability mass function of $N$ for the data $s^{(3)}$ in Fig.~\ref{fig:the_data} and the prior in eqn.~\ref{eq:prior} ($n_\text{max}=35$). 

\subparagraph{Summarizing the posterior.}
Summarizing the posterior mass function of $N$, its median is $19$ and its high-mass credible subset $\mathcal{H}_{0.2}=\{15, ..., 25\}$ (highlighted in Fig.~\ref{fig:the_posterior}). 
For what it's worth, the data in Fig.~\ref{fig:the_data} was generated from a tank population size of $n=20$ (explaining the choice of scale in Fig.~\ref{fig:the_data}). 

\subparagraph{Querying the posterior.} Suppose our military strategy would change if the size of the tank population were to exceed 30. From the posterior distribution of $N$, we calculate $\pi_{\text{posterior}}(N>30 \mid M^{(3)}=15)\approx 0.066$.

\begin{figure}[h!]
        	\centering
	\begin{subfigure}[b]{0.4\textwidth}
        		\includegraphics[width=\textwidth]{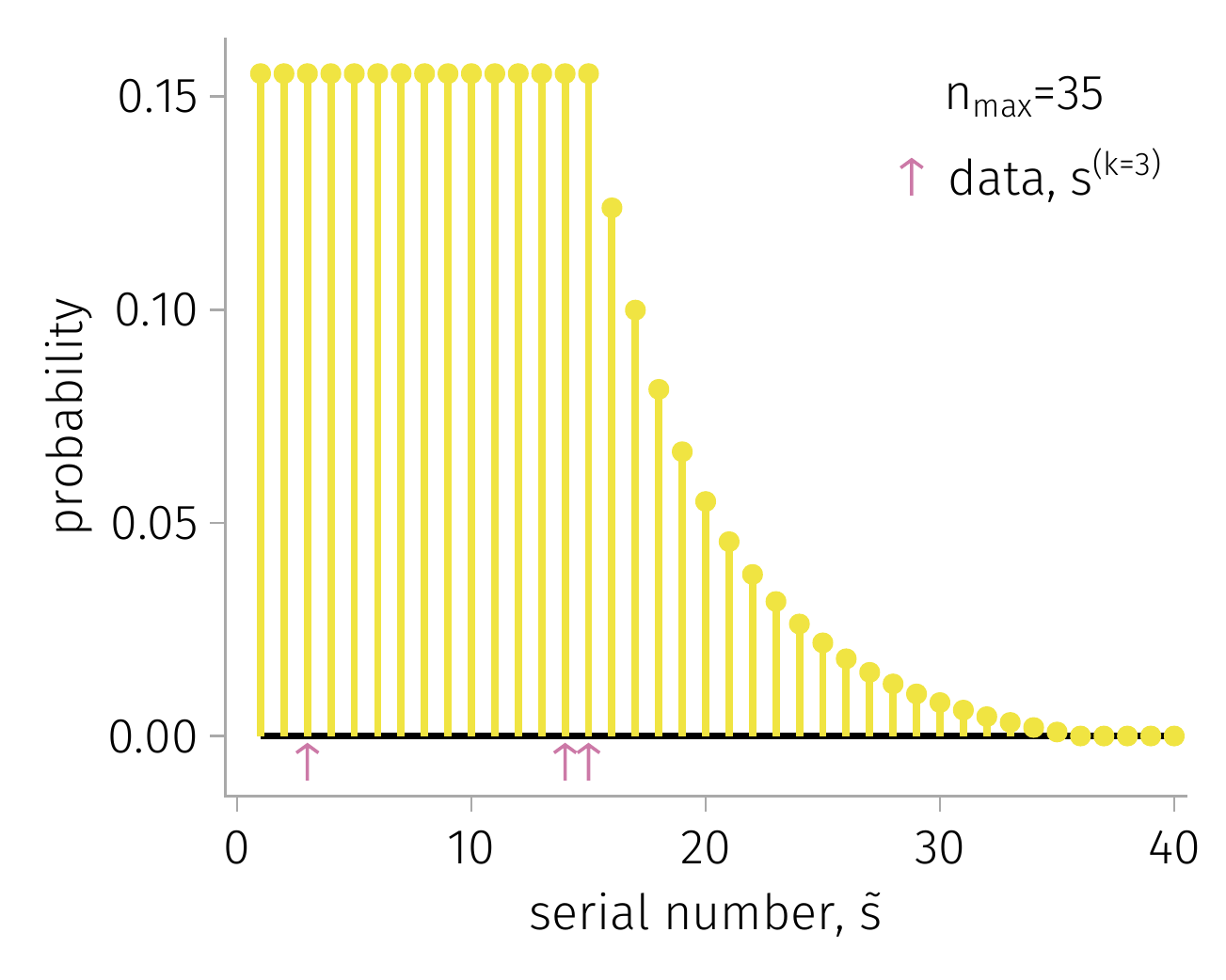}
        		\caption{posterior predictive check} \label{fig:posterior_checking}
        	\end{subfigure}
	
        	\begin{subfigure}[b]{\textwidth}
		\includegraphics[width=\textwidth]{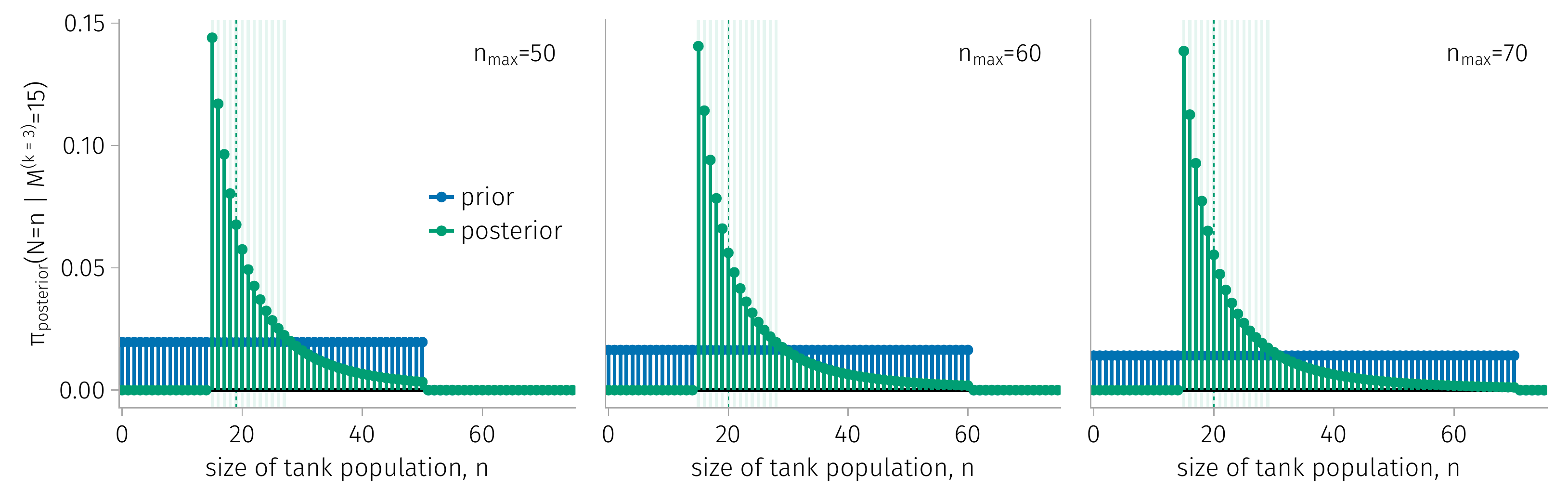}
        		\caption{sensitivity of the posterior to the prior} \label{fig:posterior_sensitivity}
        	\end{subfigure} 
            \caption{Checking (a) the consistency of the data $s^{(3)}$ with the probability of observing each serial number under the tank-capturing process and the posterior mass function of $N$ and (b) the sensitivity of the posterior mass function of $N$ to the upper bound $n_{\text{max}}$ imposed by the prior mass function of $N$.}
\end{figure}

\subparagraph{Posterior predictive checking.} As a posterior predictive check, Fig.~\ref{fig:posterior_checking} shows how the observed serial numbers in the data $s^{(3)}$ compare with the probability of observing each serial number under the posterior mass function of $N$, according to eqn.~\ref{eq:posterior_check}.

\paragraph{Sensitivity of the posterior to the prior.} Because of the subjectivity involved in constructing the prior, checking the sensitivity of the posterior to the prior is good practice \cite{van2021bayesian}. Fig.~\ref{fig:posterior_sensitivity} shows how the posterior mass function of $N$ changes as we increase the upper-bound on the tank population $n_\text{max}$ we impose via the prior mass function of $N$ in eqn.~\ref{eq:prior}. 
The median of the posterior under $n_\text{max} \in \{60, 70\}$ is 20 (an increase of one compared to $n_\text{max}=35$). The maximum of the high-mass subset $\mathcal{H}_{0.2}$ increases to 29 for $n_\text{max}=70$. 

\begin{figure}[h!]
        	\centering
	\begin{subfigure}[b]{0.5\textwidth}
        		\includegraphics[width=\textwidth]{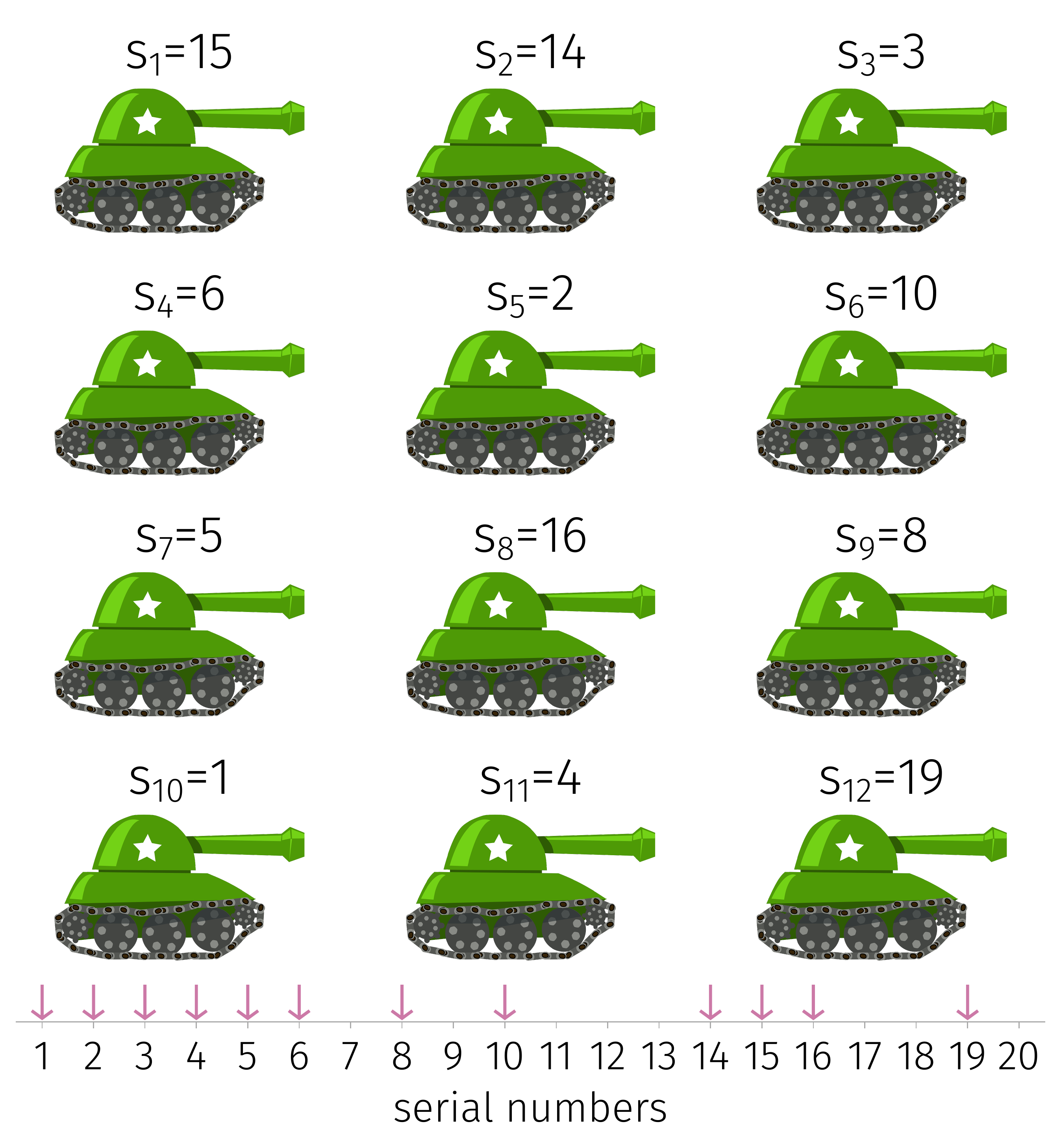}
		\caption{the updated data $s^{(k=12)}$}
	\end{subfigure}
	\begin{subfigure}[b]{0.5\textwidth}
	 \includegraphics[width=\textwidth]{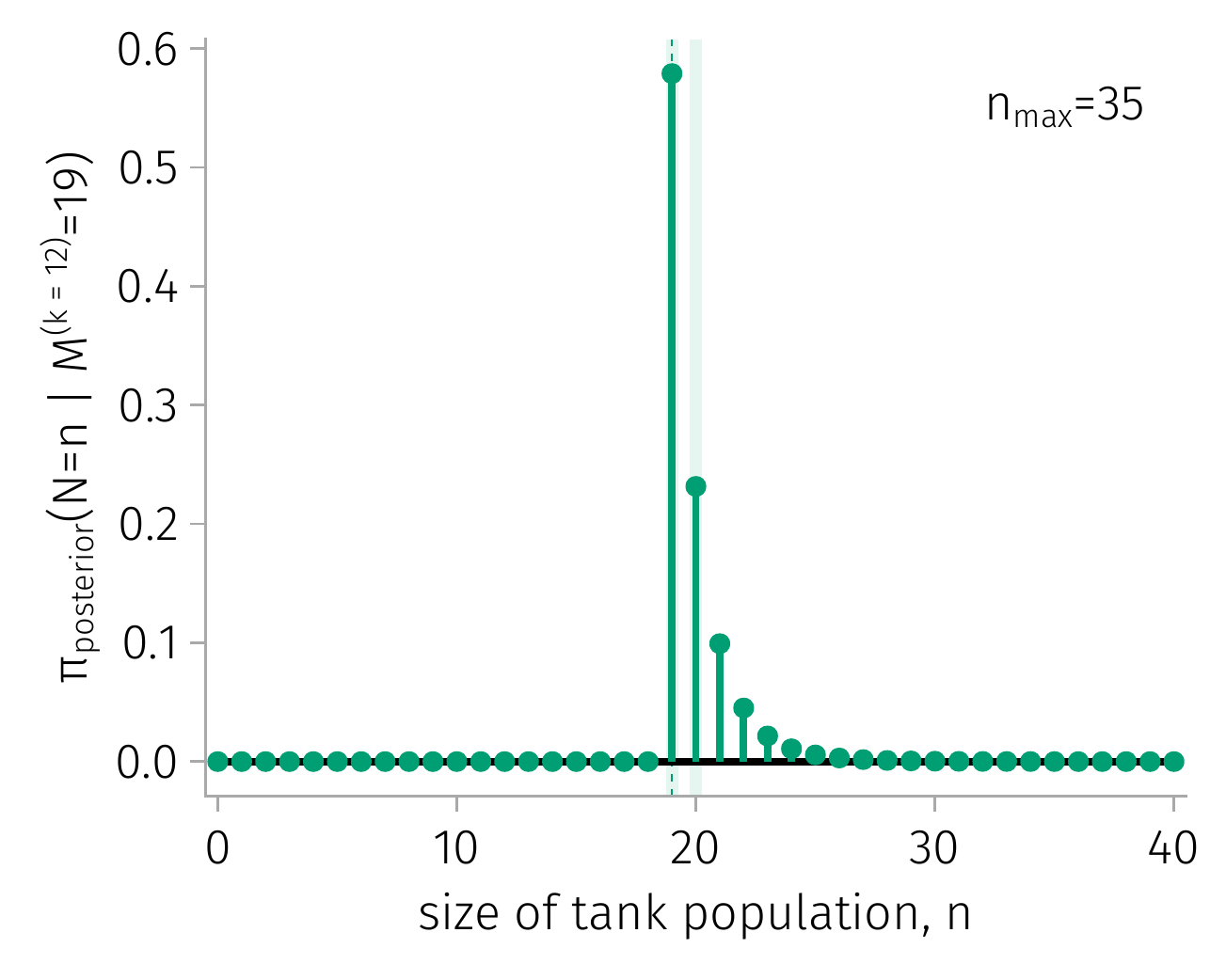}
	 \caption{the updated posterior mass function of $N$}
	     	\end{subfigure}
            \caption{The updated posterior mass function of $N$ (b) after we capture an additional 9 tanks with serial numbers in (a).
            	} \label{fig:moretanks}
\end{figure}

\paragraph{Capturing more tanks.} Suppose we capture an additional 9 tanks and re-run the Bayesian analysis. Fig.~\ref{fig:moretanks} shows the updated posterior mass function of $N$. The high-mass credible subset $\mathcal{H}_{0.2}$ shrinks considerably, to $\{19, 20\}$. This shows how more data---increasing the number of tanks captured, $k$---generally reduces our uncertainty about the tank population size.

\section{Discussion}
\paragraph{Selection bias.}
A strict assumption in the textbook-friendly German tank problem, which enables us to estimate the size of the population of tanks from a random sample of their (sequential) serial numbers, is that sampling is uniform. 
To check consistency of the sample with this model of the tank-capturing process, Goodman \cite{goodman1954some} demonstrates a test of the hypothesis that the sample of serial numbers is from a uniform distribution. 
Interesting extensions of the textbook German tank problem could involve modeling selection bias in the tank-capturing process. 
Such bias could arise eg.\ hypothetically, if older tanks with smaller serial numbers were more likely to be deployed in the fronts opened earlier in the war, where capturing tanks is more difficult than at less fortified fronts opened more recently.


\paragraph{The German tank problem in other contexts.}
The Bayesian probability theory to solve the German tank problem applies (perhaps, with modification) to many other contexts where we wish to estimate the size of some finite, hidden set \cite{cheng2020estimating}, eg.: the number of taxicabs in a city \cite{grajalez2013great}, the number of accounts at a bank \cite{hohle2006bayesian}, the number of furniture pieces purchased by a university \cite{goodman1954some}, the number of aircraft operations at an airport \cite{mott2016estimation}, the extent of leaked classified government communications \cite{gill2015estimating}, the time needed to complete a project deadline \cite{fehlmann2017new}, the time-coverage of historical records of extreme events like floods \cite{prosdocimi2018german}, 
 the length of a short-tandem repeat allele \cite{tang2017profiling}, the size of a social network \cite{katzir2011estimating}, the number of cases in court \cite{wu2022augmenting}, the lifetime of a flower of a plant \cite{pearse2017statistical}, or the duration of existence of a species \cite{roberts2003did}.
Mark and recapture methods in ecology to estimate the size of an animal population \cite{nichols1992capture,chao2001overview} are tangentially related to the German tank problem.

\paragraph{The practice of inscribing sequential serial numbers on military equipment.}
Germany adopted the practice of marking their military equipment with serial numbers and codes to trace the equipment/parts/components back to the manufacturer. However, the sequential nature of these serial numbers was exploited by the Allies to estimate their armament production. 
To reduce vulnerability to serial number analysis for estimating production while maintaining advantages of tracing equipment back to the manufacturer, serial numbers and codes could instead be obfuscated by eg.\ chaffing \cite{rivest1998chaffing}. 

\section*{Data and code availability} The Julia \cite{bezanson2012julia} code to reproduce all of our visualizations drawn using Makie.jl \cite{DanischKrumbiegel2021} is available on Github at \url{github.com/SimonEnsemble/the_German_tank_problem}.

\section*{Acknowledgements}
Thanks to Bernhard Konrad for providing detailed feedback on the first draft and to my students Gbenga Fabusola, Adrian Henle, and Paul Morris for feedback on the introduction. 

\bibliography{refs}
\bibliographystyle{unsrt}

\end{document}